\pdfoutput=1 
\documentclass{JINST}

\title{Photon Detection in the Cryogenic Apparatus for Precision Tests of Argon Interactions with Neutrinos (CAPTAIN)}

\author{K. Rielage$^a$ for the CAPTAIN Collaboration\\
\llap{$^a$}Los Alamos National Laboratory,\\
  Los Alamos, NM  87545, USA\\
E-mail: \email{rielagek@lanl.gov}}

\abstract{The Cryogenic Apparatus for Precision Tests of Argon Interactions with Neutrinos (CAPTAIN) is being built at Los Alamos National Laboratory.  A hexagonal time projection chamber (TPC) with a 1 m drift length will be constructed inside a cryostat containing 7,700L of liquid argon.  CAPTAIN will be used to test interactions using beams of neutrons and neutrinos.  It will serve as a test bed for various options for the Long Baseline Neutrino Experiment (LBNE) including in the photon detection system.  The current photon detection system will be described and future options discussed.  The system is composed of sixteen R8520-500 Hamamatsu photomultiplier tubes with a wavelength shifting coating on acrylic in front of the PMT.  Various wavelength shifting coatings can be examined with the current default of tetraphenyl butadiene.}


\begin{document}

\section{Introduction to CAPTAIN Detector}
The Long-Baseline Neutrino Experiment (LBNE) \cite{bibLBNE} has proposed to construct two 9.4 kton liquid argon far detectors at the Sandford Underground Research Facility (SURF) in Lead, SD to detect neutrinos created by a new beamline generated at Fermilab.  The ultimate goals of LBNE are to determine the neutrino mass hierarchy, precisely measure the mixing angles and mass splittings in the neutrino flavor-oscillation matrix, and search for and measure leptonic CP violation.  In addition, the large far detector could be used to search for nucleon decay and make a high-precision measurement of neutrinos from galactic supernovae bursts.  

In order to achieve these goals significant studies are needed using a liquid argon TPC.  The construction of a reasonably-sized, portable liquid argon TPC is under way at Los Alamos National Laboratory.  This detector, called CAPTAIN for Cryogenic Apparatus for Precision Tests of Argon Interactions with Neutrinos, will contain approximately 7,700 liters of liquid argon with approximately 3,400 liters contained within the TPC.  The construction effort is being completed under the Laboratory Directed Research and Development program.  The goals of the project are to gain experience with constructing and operating a liquid argon TPC, commissioning the detector with cosmic-ray muons, trying to identify captures on argon of $\mu^-$ events which will allow the possibility of $\mu^-/\mu^+$ separation for CP violation measurements, and to perform supernova-related studies such as spallation backgrounds and low energy particle identification.  In addition, the detector will serve as a test bench for calibration development such as laser calibration of the liquid argon drift constants, and for tests of a variety of photon detection systems.

\begin{figure}[tb] 
\centering
\includegraphics[width=.6\textwidth]{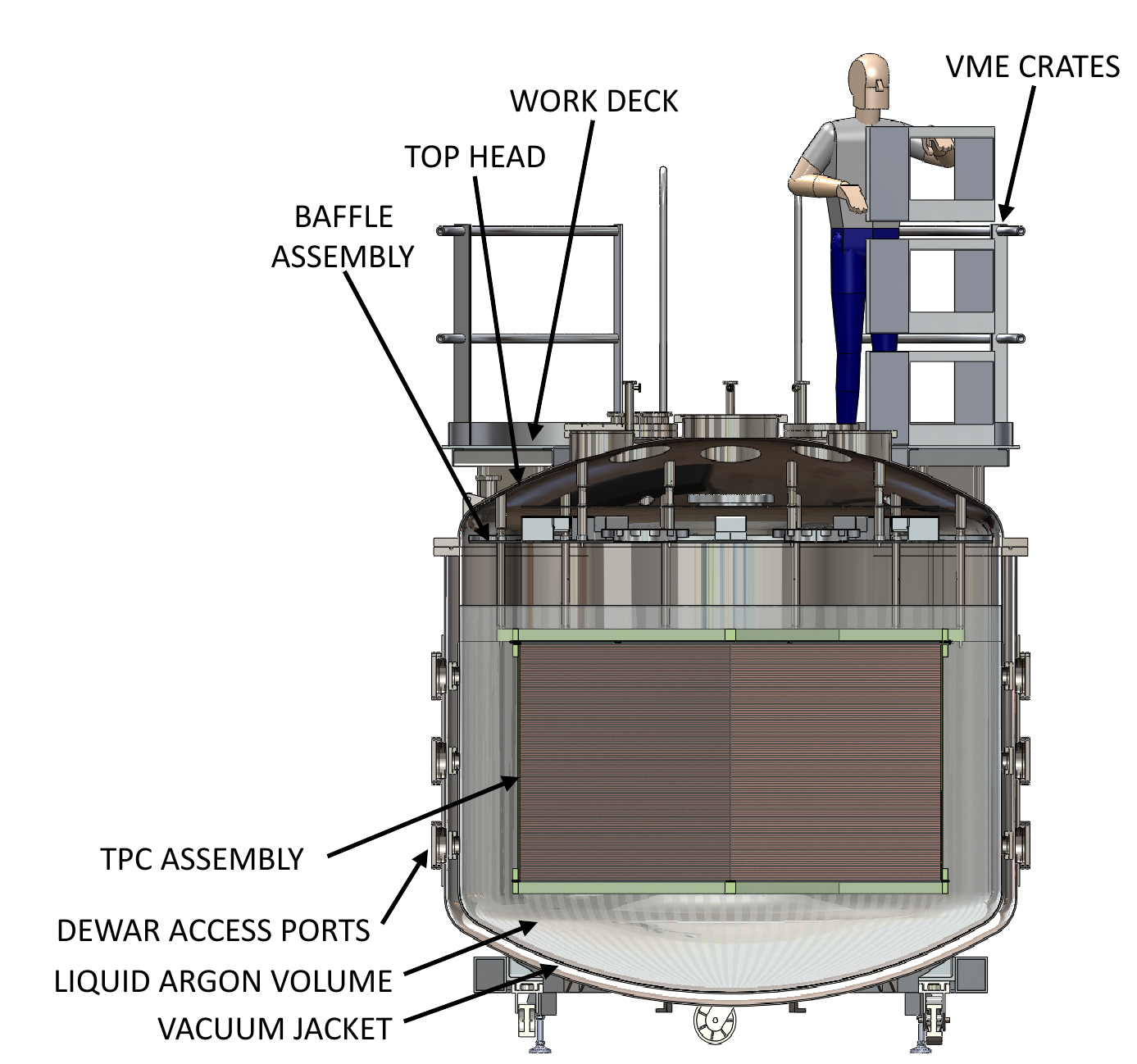}
\caption{A cutaway model of the CAPTAIN detector.  The TPC inside the cryostat is 2m long and 1m high.}
\label{fig:assembly}
\end{figure}

The full CAPTAIN detector is shown in Figure \ref{fig:assembly}.  The cryostat is a stainless steel, vacuum jacketed pressure vessel with a single large removable flange at the top.  The top flange supports the TPC and all the internal systems so that the detector can be assembled and maintained by removing the entire detector with all connections out of the cryostat by lifting the top head.  The top flange has a number of smaller flanges for electrical connections, process systems and monitoring.  Side ports on the cryostat contain several windows to allow laser light to pass through the active liquid argon volume for calibration.  The TPC has six sides, each 2 m across parallel sides with a 100 cm drift. The TPC is constructed of FR-4, a fiberglass epoxy material, and has a wire mesh cathode, two induction planes, a collection plane, a wire grid plane to define the first induction plane signal, and a ground plane for HV return.  The plane spacing is 3.18 mm and the wire spacing is 3 mm.  The drift field is 500 V/cm resulting in a drift velocity of 1.6 mm/$\mu$s.  The TPC is composed of two subassemblies, a field cage and a wire plane assembly.  The detector is utilizing the MicroBooNE electronics with a cold and a warm section.  The cold electronics are the front-end ASIC amplifiers that are in close proximity to the wire termination boards from the TPC and a cable that brings the signals from the front-end board to a cryogenic feedthrough to the warm side.  On the warm side, the signals are amplified further and driven on to the ADC module and data acquisition system.    The TPC electronics will integrate with the photon detection system and external triggers such as from a muon telescope or beamline.

The CAPTAIN detector is expected to be commissioned in late 2013/early 2014.  

\section{Planned Beam Tests}
In order to achieve the experiment's goals, several beamlines are being considered for detector running.  These include a neutron beam at the Los Alamos Neutron Science Center (LANSCE), a neutrino beam from a stopped pion source at the Spallation Neutron Source (SNS) at Oak Ridge National Laboratory, and the NuMI neutrino beam at Fermilab.  A run in a neutron beam would allow for the characterization of neutron interactions in a liquid argon TPC, to measure the response to multi-particle events in the high-energy regime, and characterize the reconstruction efficiency of these events.  The LANSCE beamline can deliver neutrons with energy up to 800 MeV.  A neutrino run at SNS would provide events with a spectrum that overlaps with the expected neutrino spectrum from a supernova.  If the detector was placed within 50 m of the SNS target several thousand events per year would be measured.  The CAPTAIN detector would contain about 10\% of the neutrino events in the NuMI beam in an on-axis location.  This would provide about 370,000 such events in a year of running.  The data could be used to measure the neutrino-Ar cross-sections above 2 GeV and provide data for event reconstruction algorithm improvements in this energy regime.

\section{Photon Detection System}
By detecting the scintillation light produced during interactions in the CAPTAIN detector, the photon detection system can provide valuable information.  Detection of several photoelectrons per MeV for a minimum ionizing particle (MIP) in a TPC with a field of 500 V/cm improves the energy resolution of the detector by 10-20\%.  The scintillation light can also be used to determine the energy of neutrons using time of flight when the experiment is placed in a neutron beamline.  

\begin{figure}[tb] 
\centering
\includegraphics[width=.4\textwidth]{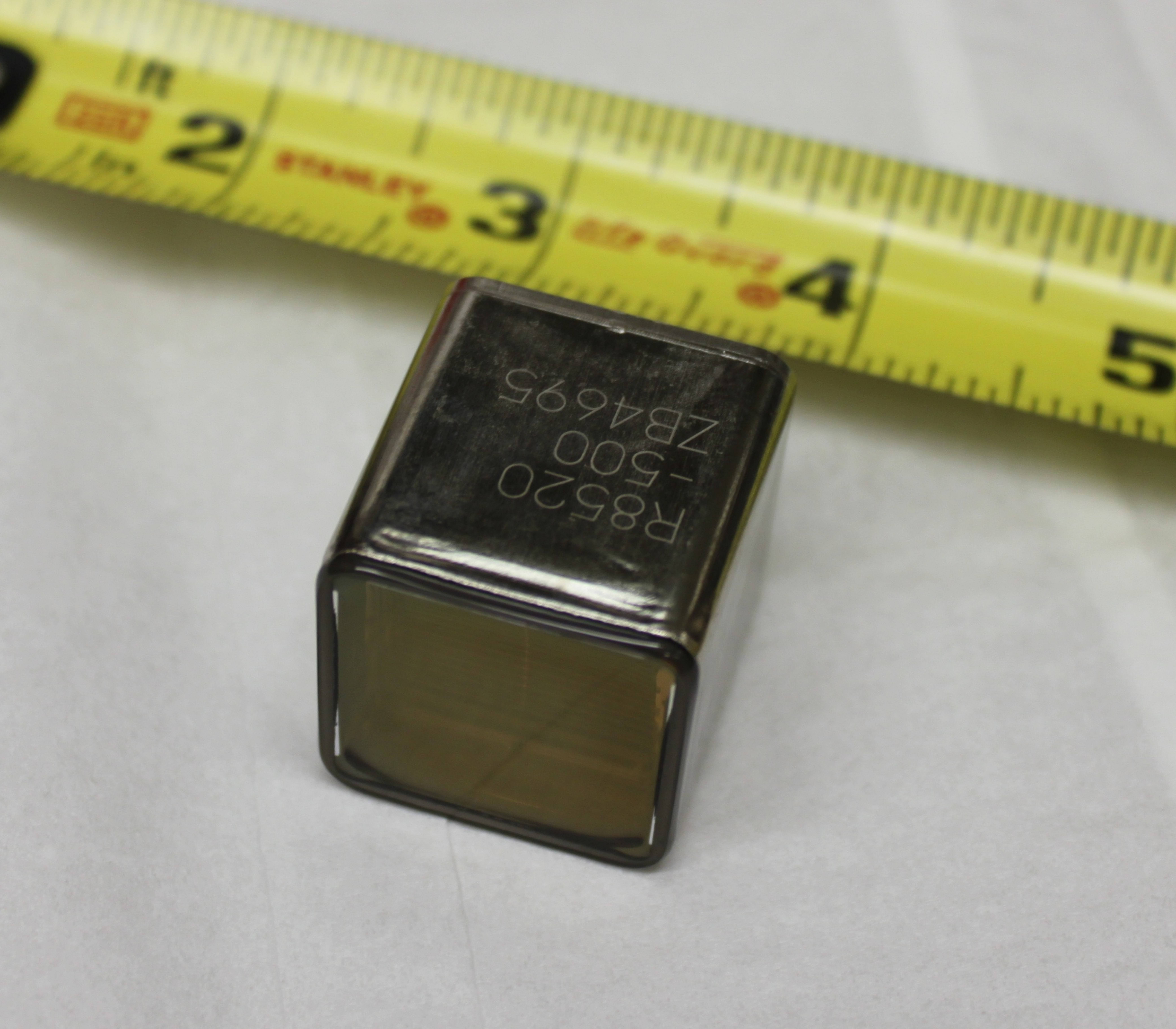}
\caption{Photo of the Hamamatsu R8520-500 photomultiplier tube.}
\label{fig:PMT}
\end{figure}

Liquid argon scintillates at a wavelength of 128 nm which unfortunately is readily absorbed by most photodetector window materials.  It is thus necessary to shift the light to the visible.  The photon detection system is composed of a wavelength shifter covering a large area of the detector and a number of photodetectors to collect the visible light.  The baseline CAPTAIN photon detection system uses tetraphenyl butadiene (TPB) as a wavelength shifter and sixteen Hamamatsu R8520-500 photomultiplier tubes (PMT) (see Figure \ref{fig:PMT}) for light detection.  The R8520 is a compact PMT approximately 1'' x 1'' x 1'' in size with a borosilicate glass window and a special bialkali photocathode capable of operation at liquid argon temperatures (87 K).  It has a 25\% quantum efficiency at 340 nm \cite{bibPMT}.  TPB is the most commonly used wavelength shifter for liquid argon detectors and has a conversion efficiency of about 120\% when evaporated in a thin film.  It has a re-emission spectrum that peaks at about 420 nm \cite{bibTPB}.  The TPB will be coated on a thin piece of acrylic in front of the PMTs.    Eight PMTs will be located on top of the TPC volume and eight on the bottom.  This will provide a minimum detection of 2.2 photoelectrons per MeV for a MIP.  The amount detected will increase if the entire top and bottom surfaces are coated with TPB.  

The PMTs will use a base with cryogenically compatible discrete components.  The cable from the base to the cryostat feedthrough is Gore CXN 3598 with a 0.045" diameter to reduce the overall heat load.  The PMT signals will be digitized at 250 MHz using two 8-channel CAEN V1720 boards.  TDC boards may be utilized to improve neutron time of flight measurements during neutron running.  The digitizers are readout through fiber optic cables by a data acquisition system written for the MiniCLEAN experiment \cite{bibGastler}.

The CAPTAIN detector will serve as a test bench for alternative photon detection systems.  Such options include testing other wavelength shifting films, acrylic light guides to the photodetectors, and other photodetectors such as SiPMs, larger PMTs capable of cryogenic operation, and avalanche photodiode arrays.

\acknowledgments

This work is supported by the Los Alamos National Laboratory's Laboratory Directed Research and Development program and the Department of Energy's Office of Science High Energy Physics program.

\end{document}